# Robustly estimating the COVID19 epidemic curve in northern Italy using all-cause mortality

Luca Presotto[1]

[1] **Nuclear Medicine Unit,** IRCCS Ospedale San Raffaele, Milano (MI) Italy

Correspondence: presotto.luca@hsr.it

**Abstract**

**Background:** Northern Italy was one of the most impacted areas by COVID. It is now widely assumed that the virus was silently spreading for at least 2 weeks before the first patient was identified. During this silent phase, and in the following weeks when the hospital system was overburdened, data collection was not performed in an accurate enough way to estimate an epidemic curve. With the aim of assessing both the dynamics of the introduction of the virus and the effectiveness of containment measures introduced, we try to reconstruct the epidemic curve using all cause mortality data.

**Methods:** we collected all cause mortality data stratified by age from the national institute of statistics, together with COVID-related deaths data released by other government structures. Using a SEIR model together with estimates of the exposure to death time distribution, we fitted the reproduction number in different phases of the spread at regional level.

**Results:** We estimate a reproduction number of 2.6±0.1 before case 1 was identified. School closures in Lombardy lowered it to 1.3. Soft lockdown measures resulted in R<0.8 and no further reductions were observed when a hard lockdown was introduced (e.g. Emilia-Romagna soft lockdown 0.67 ±0.07, hard lockdown 0.69±0.071). Reproduction number for the >75 age range during hard lockdown are consistently higher than for the rest of the population (e.g. 0.98 vs 0.71 in Milan province), suggesting outbreaks in retirement facilities. Reproduction numbers in Bergamo and Brescia provinces starting from March 7[th] are markedly lower than in other areas with the same strict lockdown measures (Nearby provinces: 0.73, Brescia: 0.52, Bergamo 0.43) supporting the hypothesis that in those provinces a large percentage of the population had already been infected by the beginning of March.

# Introduction

Northern Italy is one of the regions of the world where the COVID-19 pandemic stroke hardest [1, 2]. Specifically, in the province of Bergamo, the excess mortality compared to the previous 5 years was of 4,500 people in the month of march alone, 0.4% of the population [3]. As the current best estimates for the infection fatality rate (IFR) are in a range between 0.5% and 1% [4, 5], it indicates a very large spread in the province.

Data collection, especially in the early phases of the pandemics, was incomplete, it did not involve contact tracing and only highly symptomatic subjects were tested. This can be seen, as an example, from the official case fatality rate (CFR) for Lombardia: it is 35% for all cases that had an outcome before April 30[th] [6]. Due to a collapse of the healthcare systems[7, 8], even subjects with severe symptoms were not being tested at the peak of the peak and out of the 6,100 extra deaths in the Bergamo province, only 2,500 are officially attributed to COVID [9].

The analysis of the epidemic curve ("epi-curve") is important to understand how this situation materialized. As Italy underwent one of the strictest and longest lockdowns, it is also important to assess which measures were the most effective in reducing the spread [10].

There are no public COVID data which are robust or complete enough to estimate the reproduction number of this virus in Italy over time. We therefore decided to try to model the reproduction number in different phases of the epidemics using all-cause mortality data from the Italian national institute for statistics (ISTAT).

# Materials and Methods

**Data**

All cause mortality data, stratified by province and by age, were downloaded from the ISTAT website. Data claim to cover more than 99% of the Italian population and are provided until May the 31[st] 2020 [11]. They include also data from the previous 5 years, that can be used to establish a baseline.

We used a second dataset provided by the Civil Protection mechanism of COVID certified deaths [12]. These data however are not stratified by age, deaths are counted according to the date of notification and not by date of death (providing large inconsistencies in some cases). In regions that were most impacted (northern Italy) such data do not explain up to 50% of the excess deaths seen in the previous dataset [9].

**Mathematical model**

We describe the number of infected people using an SEIR compartmental model (Susceptible-Exposed-Infectious-Recovered) according to the following differential equation

$$\begin{cases} \dfrac{dS}{dt} = -R(t)\dfrac{I}{\tau_i} \\ \dfrac{dE}{dt} = R(t)\dfrac{I}{\tau_i} - \dfrac{E}{\tau_e} \\ \dfrac{dI}{dt} = \dfrac{E}{\tau_e} - \dfrac{I}{\tau_i} \end{cases}$$

Where $S$ represent the number of susceptible individuals, $E$ the number of individuals that have been exposed but are not infective yet, $I$ the number of currently infective individuals. $R(t)$ is the reproduction number at a specific day, and we assume this function to be piecewise constant, with changes when new restrictions were implemented. The time constants for the transition from exposed to susceptible and from infective to isolated were set to $\tau_i = 4d$ and $\tau_e = 1d$ according to the literature to reproduce the observed temporal distributions with a mean exposure to onset of about 5 days [13, 14]. We model the number of deaths starting from the number of subjects that transition from the susceptible to the exposed one in a day. We then convolve this distribution with the expected time distribution of exposure-to-death. This model does not account for reductions in the number of susceptible individuals, which result in a lower $R(t)$.

*Model fit*

The SEIR model was integrated numerically (using steps of 0.002 d) and from its results an expected number of exposed subjects per day was measured. This allowed fitting the observed number of excess deaths using a Poisson model: $\ell = \sum_i \bar{y}_i - y_i \ln \bar{y}_i$, with $\bar{y} = D(R(t), I_0) + baseline$. As previously stated, we impose R(t) to be piecewise constant, with 3 total changes, for a total of 4 R values to be fit, on top of $I_0$, the number of infectious subjects at day 0 (conventionally positioned at February the 1st.)

After the fit, we inverted the hessian to estimate both the error on the fitted parameters and the correlation between them. When the estimation was unstable, typically for the first R, the fit was repeated with R fixed.

*Baseline estimation*

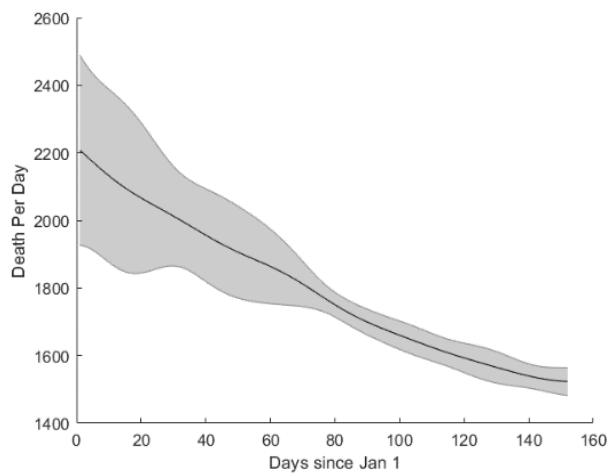

The baseline was estimated as the mean of deaths per day over the age range under analysis for all of Italy in the previous 5 years, then filtered with a zero-phase 4th order Butterworth filter with a period of 20 days. The curve was then scaled to the fraction of deaths of the region under analysis compared to all of Italy. Due to the influence of the flue season, we find the coefficient of variation between the previous years to be higher than 12% in January, and it gradually decreases to 3% from the beginning of April. The estimated baseline is shown in Figure 1.

*Figure 1 Estimated baseline of deaths per day in the whole of Italy. Shaded area represents 68% confidence level estimated from previous 5 years*

*Exposure to death distribution*

The exposure to death distribution, central to this paper, was built using data reported by Linton et al[13], by convolving the reported exposure to onset and onset to death distributions. Specifically, we used a gamma function with mean 6.0 days and 3.1 days SD for incubation period and a gamma function with mean 15.0 days and 6.9 days SD.

**Variations in reproduction number**

A summary of the lockdown measures introduced in Italy can be found here[15]. For provinces of Lombardy, we considered 4 intervals for the estimation of the reproduction number R:

1. Until February 23rd, were the first closures were announced. The first patient ever in Italy was identified February the 21st.
2. February the 24th till March the 7th, where the population had higher awareness, restaurants and bars were still open, even if pubs, clubs and schools were not, and there were no restriction on productive activities.
3. March the 8th till March 21st, Lombardy was put into "soft lockdown", with public exercises closed, meetings between people forbidden but productive activities still allowed, even if remote work was strongly encouraged.

4. March 22nd onwards: "hard lockdown". It was illegal to go outside of everyone's place of living for any reasons with the exclusion of a very short list of essential reasons, which did not allow even physical activity. An extremely narrowly defined series of "essential activities" were allowed to remain open.

Outside of Lombardy, different restrictions were introduced therefore we chose different time intervals. Namely

1. Until February 23rd, when we can assume that behaviours were not influenced by the pandemic as the presence of the virus was unknown.
2. February 24th till March 11th. Outside of Lombardy schools and almost all activities remained open, even if some minor restrictions were applied and it can be reasonably presumed that the population changed their behaviour, due to the news of the epidemic spread.
3. March 12th till March 21st : initial "soft lockdown" (same rules as in Lombardy)
4. March 22nd onwards: "hard lockdown" (same rules as in Lombardy)

The fit often proved to be difficult/impossible for the period of time 1 (before February 23rd) therefore it was estimated only in the provinces of Bergamo and Brescia and for age ranges >65; it was fixed to 2.6 elsewhere. For fits of data of areas outside Lombardy, where population level circulation started much later and therefore modelling period "1" is largely irrelevant, we assumed R in period "1" to be identical to that in period "2" and reduce the number of parameters.

**Settings analyzed**

We fitted the epi-curves from all-cause mortality data to the provinces of Bergamo and Brescia stratified by age (<65, 65-74,>74). We also fitted the data summing the provinces of Milano with "Monza-Brianza" only in the 65-74 and >74 age ranges, as too few deaths were recorded in the <65 range. We also performed fits to data from all of Emilia-Romagna and all of Piemonte, still only in the 65-75 and >74 range, due to the low number of deaths in these regions. In the Veneto region, we fitted data at the regional level only in the >74 range. No other region had an increase of deaths such that it was possible to perform a reliable fit.

The curve was also fit to certified COVID19 deaths reported from "Protezione Civile", in the whole central and southern Italy (excluding Sicilia and Sardegna, that, as islands, might have a different dynamics). These data are expected to be complete, as very few cases very registered in these regions.

**Estimating confidence interval of fitted parameters**

Two kind of uncertainties affect the estimated parameters: Poisson noise in the recorded number of deaths and systematic errors in the parameters of the model. The error due to Poisson noise is estimated by inverting the Hessian of the likelihood. Errors are reported as 1 standard deviation.

*Systematical sources of uncertainty*

The model is influenced by the estimate of the baseline and by the estimate of the exposure-to-death time distribution.

The estimate of the baseline has a particular impact on the fitting of the initial slope of the curve, and it severely impacts the possibility to fit data when the number of deaths over the baseline is small (e.g.: <20% increase) .

The exposure-to-death distribution also has a major impact on the fit quality. Its exact shape is largely irrelevant in points where the curve is smooth, therefore allowing a robust fit of the parameters describing the central part of the epi-curve, but it influences vastly the fit at the beginning and at the end of the outbreak.

**Estimation of the epi-curve in the province of Bergamo**

To estimate when the virus started having widespread population circulation in the province of Bergamo, we try to estimate the date in which the 200[th] person was infected, the number of new exposures per day on the 20[th] of February (the day in which the first ever case in Italy was identified), and the fraction of population exposed in 3 relevant dates (February 23[rd], March 7[th] and March 21[st] ). Such information depend on the doubling time as determined by the SEIR model (dependent on R before the 23[rd] of February), which proved difficult to fit, and on the infected fatality rate (IFR) for the whole population, which is a free parameter that allows the conversion from the estimated number of infected people to the estimated number of deaths. As these parameters have large uncertainties, we perform these computations assuming 3 different scenarios:

|  | Scenario A | Scenario B | Scenario C |
| --- | --- | --- | --- |
| IFR | 2.2% | 1.5% | 1% |
| R | 3 | 2.6 | 2.2 |
| Doubling time (d) | 2.17 | 2.62 | 3.36 |

It should be noted that scenario A is the one that estimates the latest introduction of the virus while scenario C the earliest one. Scenario B represent our best estimate.

# Results

The estimated number of deaths are reported in the following tables. Examples of data and fitted models are shown in Figure 2. In Figure 3 we show the different curves of different age ranges in the province of Milano and Monza-Brianza.

*Table 1: Deaths in Bergamo and Brescia*

|  | <65 | | | 65-74 | | | >75 | |
| --- | --- | --- | --- | --- | --- | --- | --- | --- |
|  | BG | BS | BG+BS | BG | BS | BG+BS | BG | BS |
| 1/2 – 22/2 | 2.6† | 2.6† | 3.4± 1.1 | 3.1± 0.3 | 3.3± 0.8 | 2.8± 0.2 | 2.46± 0.07 | 3.1± 0.6 |
| 23/2 - 6/3 | 1.27 ± 0.07 | 1.46± 0.11 | 1.18± 0.05 | 1.16± 0.03 | 1.35± 0.04 | 1.14± 0.02 | 1.08± 0.01 | 1.39± 0.02 |
| 7/3 - 21/3 | 0.47 ± 0.06 | 0.39± 0.12 | 0.51± 0.04 | 0.46± 0.03 | 0.55± 0.03 | 0.51± 0.02 | 0.50± 0.01 | 0.68± 0.01 |
| 22/3 - 12/4 | * | * | 0.79± 0.08 | 0.62± 0.08 | 0.69± 0.07 | 0.69± 0.05 | 0.53± 0.04 | 0.67± 0.02 |

*Fit not converging/not robust due to too few predicted deaths. Fits nonetheless exclude an expanding scenario;
† Parameter fixed and not fitted

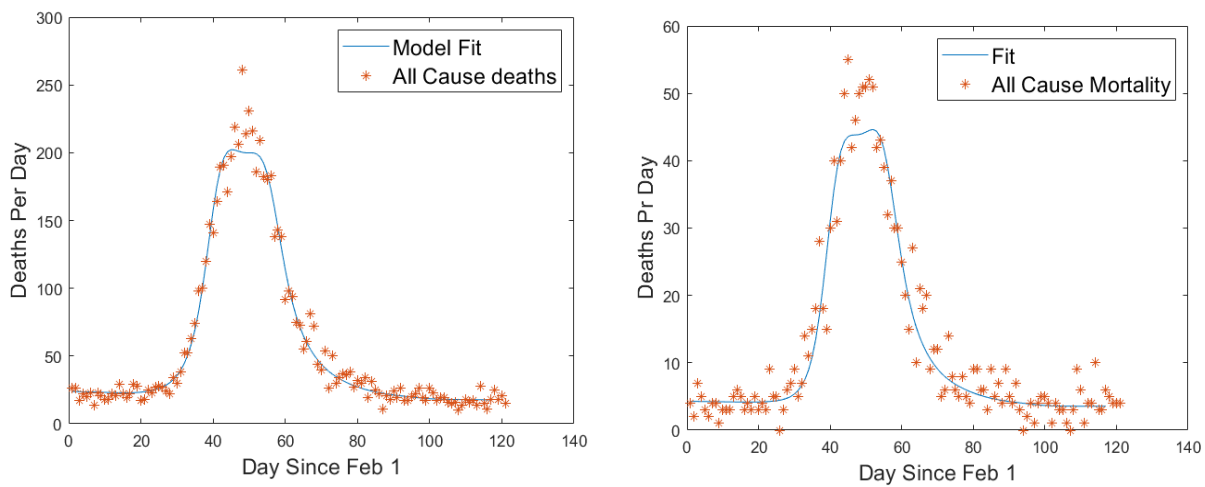

*Figure 2 Recorded all cause mortality deaths in the province of Bergamo for people over 74 years of age (left) and between 65 and 74 (right).*

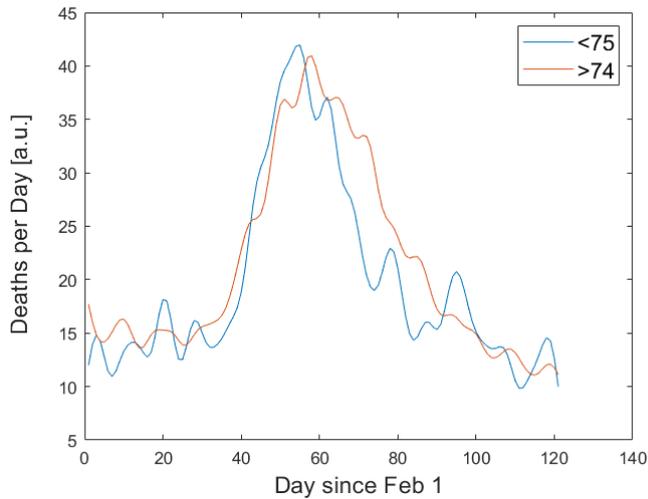

*Figure 3 Total deaths in different age ranges in the Provinces of Milano and Monza-Brianza. Data smoothed with a 6 days cutoff. The slower increase and much slower decrease can be easily noticed. The two curves have been arbitrarily scaled for easier visual comparison.*

*Table 2:* Deaths in Milano and Monza-Brianza

|  | 65-75 | >75 |
|---|---|---|
| 23/2 - 6/3 | 1.61 ± 0.07 | 1.44± 0.03 |
| 7/3 - 21/3 | 0.73 ± 0.03 | 0.96± 0.01 |
| 22/3 - 12/4 | 0.63 ± 0.07 | 0.73± 0.01 |

*Table 3:* Deaths in Emilia-Romagna

|  | 65-75 | >75 |
|---|---|---|
| 1/2 – 29/2 | 1.83± 0.10 | 1.88±0.06 |
| 1/3 - 11/3 | 0.87 ± 0.05 | 0.92± 0.03 |
| 12/3 - 21/3 | 0.67 ± 0.07 | 0.85± 0.03 |
| 22/3 - 12/4 | 0.69 ± 0.07 | 0.83± 0.02 |

*Table 4:* Deaths in Piemonte

|  | 65-75 | >75 |
|---|---|---|
| 1/2 – 29/2 | 1.95* | 1.95±0.17 |
| 1/3 - 11/3 | 1.22 ± 0.10 | 1.40± 0.05 |
| 12/3 - 21/3 | 0.84 ± 0.11 | 0.85± 0.03 |
| 22/3 - 12/4 | 0.62 ± 0.10 | 0.87± 0.02 |

*: Too few predicted deaths to allow fitting in this curve region. The same value fitted in the >75 age range was fixed

*Table 5:* Southern Italy

|  | Whole Population |
|---|---|
| 1/2 – 29/2 | 2.7±0.8 |
| 1/3 - 11/3 | 1.33± 0.09 |
| 12/3 - 21/3 | 0.90± 0.03 |
| 22/3 - 12/4 | 0.87± 0.01 |

*Timing of initial spread in the Bergamo province*

Our best estimate finds that on February 2nd 200 people had already been exposed in the province of Bergamo and, on the day (Feb. 20th) when the first Italian case was reported, the province of Bergamo was recording not less than 10,000 new exposures per day. Results obtained using different scenarios are reported in table 6.

*Table 6: Estimation of initial spread in the province of Bergamo*

|  | Scenario A | Scenario B | Scenario C |
| --- | --- | --- | --- |
| Day with 200th case | Feb 5th | Feb 2nd | Jan 30th |
| New exposures per day Feb 20th | 10,000 | 12,000 | 17,000 |
| Population infected Feb 21st | 2.7% | 10% | 8% |
| Population infected March 7th | 20% | 30% | 46% |
| Population infected March 21st | 24% | 37% | 56% |

**Estimation of the percentage of infected people in Bergamo and Brescia**

Using data from Piemonte and Emilia-Romagna we estimate the basic reproduction number during the "soft" lockdown to be 0.75±0.10 for the 65-74 age range. Compared to 0.46±0.03 in Bergamo, it suggests that 43±6 % of the population was already exposed by the mid of the soft lockdown period (March 15), pointing to an IFR between 1 and 1.5%. Applying the same reasoning to the province of Brescia we estimate that 27±4% of the population was already exposed on March 15. Similar ratios are observed in the >74 age range.

# Discussion

Reconstructing epi-curves from excess death curves proved a challenging task. Due to the long delay between exposure and onset, the small percentage of individuals that die and the large spread of the exposure to death distribution, fast dynamics cannot be reconstructed. Nonetheless, a number of robust results can be extracted. One of the first limitations, is due to the fact that we can reconstruct the spread of the epidemy only for very oldest age range (>74 years in general, 65-74 in areas with large spread). As also this work shows, it cannot be assumed that the spread is uniform across all age ranges, due to the different number of contacts per day that each person has in different steps of life. Uniform spread across different age groups can be reasonably assumed in areas with very large impact, but it is not possible to exclude a much larger spread for the oldest age class during the hard-lockdown, as these people were the most likely to be in assisted living.

For this reason, we call for public authorities to release more data. Using daily hospital admission data stratified by age, which have higher numbers, much shorted delays from onset, and no baseline (which introduces both statistical noise and systematic uncertainties), higher quality curves could be reconstructed. The same could b done using the number of daily admissions to ICUs.

*Difference in epidemic curves between with age*

In all regions where the fit was possible it was found that the epi-curve had similar slopes in the initial phase for people under 75 years of age and for older people, but markedly slower decrease. Different phenomena can contribute to modify the distribution in this age range:

1. Outbreaks in retirement homes/hospitals, which were not reduced by lockdowns
2. Preventable deaths unrelated to COVID that could not be treated due to hospital collapse
3. Different infection-to-death distribution, with longer right tails
4. Decrease in the baseline due to less deaths attributable to diseases that the lockdown prevented

While most likely all of these factors contribute to modifying the distribution, the fact that in Bergamo and in Brescia, the earliest hit provinces, the estimated reproduction numbers are similar across different age ranges, hints that factor 2,3 4 have a minor impact. It should be noted that factor 4 has the effect of increasing fitted R, instead of decreasing it like factors 2 and 3.

*Difference in reproduction number between "soft" and "hard" lockdown*

In no area analysed a statistically significant reproduction number was observed between the "soft" and "hard" lockdown periods. A very small reduction was observed in the Milano area which was not statistically significant. Also, assuming an IFR of 1.0 %, we can assume by March 21st a reduction in the number susceptible individuals of ~15%, which would entirely explain the observed reduction. This points to the fact that the additional restrictions imposed by the hard-lockdown (restriction of outdoor activities and closure of all workplaces) were not effective, pointing probably the residual diffusion to be happening at homes and in essential services and to the need of quarantining infected people separated from their families and to actively trace contacts. An interesting aspect of this finding is not that the soft-lockdown was sufficient to reduce $R \approx 0.75$, but more than the hard-lockdown could not lower this value anymore. This is consistent with what was theorized initially[16], where the major impact of household transmission was suspected. Data from Germany, where lockdown measures were much softer (Oxford government response stringency index at peak 76 for Germany vs 91 for Italy[17]), show $R \approx 0.8$[18] as the minimal value over the month where the strictest measures were imposed [19]. This seems to indicate that this value is the lower limit, most likely due to household transmission, that cannot be further reduced without introducing quarantines in separated structures for family members and widespread testing.

*Initial undetected spread*

Our model shows that the virus was already spreading widely at the community level in the province of Bergamo not later than the 5th of February, 15 days before the first case was detected in Italy. At the same time, we can exclude with confidence widespread circulation before January 30th.

*Initial doubling time and reproduction number estimation*

The reproduction number before February the 23rd is the most difficult parameter to fit from this data. By pooling together results from provinces of Bergamo and Brescia over different age ranges, we obtain a best estimate of 2.6±0.1, which translates in a doubling time of 2.6 days. This confirms previous findings that described the hardest challenge while addressing the COVID-19 pandemic in the very short doubling time compared to the long delay for the effectiveness of the containment measures [20]. Such an assumption describes well the results of this paper with a fast reduction in reproduction number in Lombardy already around the end of February, but the peak of the deaths near the end of march.

*Estimated number of infected individuals in the province of Bergamo*

Comparison of the reproduction rate between multiple regions estimated that 43±6% of the population of the province was exposed by mid-March. Recently, Poletti et. Al, estimated the IFR in Lombardia for different age range, using data independent from ours [21]. Specifically, they estimate 10% IFR for the >69 age range (95% C.I.: 8.0 – 13.6). We can find 5,250 excess deaths in the province of Bergamo until May 15th, for the the >69 y age range, corresponding to 3.0% of the population of the province in this age range [95% C.I.: 2.8% - 3.2%]. This would indicate that 33% (95% C.I.: 20%-40%) of the population in this age range was infected over the whole period, which is compatible, even if lower, than our estimate.

**Study limitations**

Further studies are needed to assess the robustness of the model that we reconstruct. Using different values for $\tau_i$ and $\tau_e$ would result in changes of the estimated values of R, but it would lead to undistinguishable curves characterized by identical doubling times in the different time intervals. Also, as $\tau_i$ represents the time to subject isolation, this parameter changes over time (e.g.: with more awareness people that are currently experiencing very mild symptoms can choose to isolate anyway, contact tracing can lower $\tau_i$, and finally lockdowns might leave as the only kind of transmission at-home transmission between family members, that cannot isolate, therefore increasing $\tau_i$.) The same issue is present when estimating R from serial intervals[20]. Another issue is the unknown exposure to death curve, which current best estimate has large degrees of uncertainty. We have not modelled its impact on the fitted parameters.

Finally, modelling the baseline correctly might be crucial for regions with relatively low number of cases. This is further complicated by the fact that lockdown might have highly reduced the number of deaths due to other respiratory illness.

**Biblioraphy**